%% file: balanced.tex
\documentclass[11pt,letterpaper]{article}
\input{macros.tex}
\input{bib_macros}

\usepackage{url}

\def\cqedsymbol{\ifmmode$\lrcorner$\else{\unskip\nobreak\hfil
\penalty50\hskip1em\null\nobreak\hfil$\lrcorner$
\parfillskip=0pt\finalhyphendemerits=0\endgraf}\fi}

\usepackage{parskip}
\usepackage{microtype}
\begin{document}

\begin{titlepage}
\def\thepage{}
\thispagestyle{empty}

\title{Routing under Balance}

\date{}
\author{
Alina Ene\thanks{Department of Computer Science and DIMAP, University of Warwick, {\tt A.Ene@dcs.warwick.ac.uk}.}
\and
Gary Miller\thanks{Carnegie Mellon University, {\tt glmiller@cs.cmu.edu}.}
\and
Jakub Pachocki\thanks{Carnegie Mellon University, {\tt pachocki@cs.cmu.edu}.}
\and
Aaron Sidford\thanks{Microsoft Research New England, {\tt asid@microsoft.com}.}
}

\maketitle

\input{abstract}

\end{titlepage}

\input{intro}
\input{overview}
\input{routing}

\input{lower}
\input{maxflow}

\input{proximal}

\subsection*{Acknowledgments}
We thank Yin-Tat Lee for several helpful discussions, and in particular for his help with the results in Section~\ref{sec:maxflow}. 
This work was partially supported by NSF awards 0843915, 1065106 and 1111109, NSF Graduate Research Fellowship (grant no. 1122374) and Sansom Graduate Fellowship in Computer Science. Part of this work was done while authors were visiting the Simons Institute for the Theory of Computing, UC Berkeley.
%\newpage
\bibliographystyle{alpha}
\bibliography{balanced}

\begin{appendix}
\input{overview_proofs}
\input{ddecomp_proofs}
\input{lowstretch_proofs}
\input{racke_proofs}
\input{lower_proofs}
\input{maxflow_proofs}
\input{proximal_proofs}
\end{appendix}

\end{document}

%% file: macros.tex
\usepackage{thmtools}
\usepackage{thm-restate}

\usepackage{mathrsfs}           % \mathscr
\usepackage{vmargin,fancyhdr}   % margins
\usepackage{enumerate}

\usepackage{amsmath,amssymb}    % AMS
\usepackage{verbatim}           % comment environment
\usepackage{xspace}             % optional space
\usepackage{graphicx,float}     % figures
\usepackage{ifthen,calc}        % macros
\usepackage{textcomp}           % \textcent
\usepackage{fancybox}           % \Ovalbox
\usepackage{hhline}             % \hhline
\usepackage{float}              % \float

\usepackage[rflt]{floatflt}     % default is vflt
\usepackage[small,compact]{titlesec}
\usepackage{setspace}
\usepackage{subfigure}

\usepackage{color}
\definecolor{ForestGreen}{rgb}{0.1333,0.5451,0.1333}
\usepackage{thumbpdf}
\usepackage[letterpaper,
            colorlinks,linkcolor=ForestGreen,citecolor=ForestGreen,
            backref,
            bookmarks,bookmarksopen,bookmarksnumbered]
           {hyperref}

\usepackage{cleveref}

\setpapersize{USletter}
\setmarginsrb{1in}{.5in}        % left, top
             {1in}{.5in}        % right, bottom
             {.25in}{.25in}     % headheight, headsep
             {.25in}{.5in}      % footheight, footskip
\newcommand{\showccc}[0]{0}
\newcommand{\ccc}[2][nothing]{% comment
  \ifthenelse{\showccc=0}{}{
    \ensuremath{^{\Lsh\Rsh}}\marginpar{\raggedright\tiny\textsf{%
        \ifthenelse{\equal{#1}{nothing}}{}{\textbf{#1}\\}#2}}}}

\newcounter{hours}\newcounter{minutes}
\newcommand{\hhmm}{%
  \setcounter{hours}{\time/60}%
  \setcounter{minutes}{\time-\value{hours}*60}%
  \ifthenelse{\value{hours}<10}{0}{}\thehours:%
  \ifthenelse{\value{minutes}<10}{0}{}\theminutes}
\ifthenelse{\showccc=0}{\rhead{}}{\rhead{\today \ [\hhmm]}}
\newtheorem{theorem}{Theorem}[section]

\newtheorem{definition}[theorem]{Definition}

\newtheorem{lemma}[theorem]{Lemma}
\newtheorem{fact}[theorem]{Fact}

\newcommand{\Proof}[0]{\smallskip\noindent\textit{\textbf{Proof:}}\quad}

\newcommand{\Proofof}[1]{\smallskip\noindent\textit{\textbf{Proof of #1:}}\quad}

\newcommand{\Oh}{\ensuremath{\mathcal{O}}}
\newcommand{\Oht}{\ensuremath{\widetilde{\mathcal{O}}}}

\newcommand{\bv}[1]{\mathbf{#1}}

\newcommand{\QED}[0]{\hfill\ensuremath{\blacksquare}\medspace\\}

\newcommand{\etal}[0]{\emph{et~al.}\xspace}

\newcommand{\R}[0]{\ensuremath{\mathbb{R}}}

\floatstyle{ruled}
\newfloat{algo}{tbp}{lop}%[section]
\floatname{algo}{Algorithm}

\newcommand{\norm}[1]{\|#1\|}

\usepackage{float}
\floatstyle{plain}\newfloat{myfig}{t}{figs}[section]
\floatname{myfig}{\textsc{Figure}}
\floatstyle{plain}\newfloat{myalg}{H}{algs}[section]
\floatname{myalg}{}
\newcommand{\poly}[1]{\text{poly}(#1)}

\global\long\def\R{\mathbb{R}}

\global\long\def\boldVar#1{\mathbf{#1}}
\global\long\def\mvar#1{\boldVar{#1}}
\global\long\def\vvar#1{\vec{#1}}
\global\long\def\defeq{\stackrel{\mathrm{{\scriptscriptstyle def}}}{=}}

\global\long\def\gradient{\bigtriangledown}
 \global\long\def\grad{\gradient}

\global\long\def\innerProduct#1#2{\big\langle#1 , #2 \big\rangle}
 \global\long\def\norm#1{\big\|#1\big\|}
 
 \global\long\def\normInf#1{\norm{#1}_{\infty}}

 \global\long\def\vd{\vvar d}

 \global\long\def\vx{\vvar x}
 \global\long\def\vy{\vvar y}

 \global\long\def\mb{\mvar B}
 \global\long\def\mc{\mvar C}

 \global\long\def\mr{\mvar R}

\global\long\def\mdiag{\mvar{\texttt{d}iag}}

\global\long\def\argmin{\mathrm{argmin}}

\global\long\def\vol{\mathrm{vol}}

\global\long\def\lmax{\mathrm{lmax}}

\global\long\def\bal{\mathrm{bal}}

\global\long\def\abal{\mathrm{abal}}

\usepackage{tikz}
\usetikzlibrary{positioning,chains,fit,shapes,calc}

%% file: bib_macros.tex
  \usepackage{nth}
  \usepackage{intcalc}

%% file: abstract.tex
\abstract{
    We introduce the notion of \emph{balance} for directed graphs:
     a~weighted directed graph is $\alpha$-balanced if for every cut $S \subseteq V$, the total weight of edges going from $S$ to $V\setminus S$ is within factor $\alpha$ of the total weight of edges going from $V\setminus S$ to $S$. Several important families of graphs are nearly balanced, in particular, Eulerian graphs (with $\alpha = 1$) and residual graphs of $(1+\epsilon)$-approximate undirected maximum flows (with $\alpha=\Oh(1/\epsilon)$).

  We use the notion of balance to give a more fine-grained understanding of several well-studied routing questions that are considerably harder in directed graphs. We first revisit oblivious routings in directed graphs. Our main algorithmic result is an oblivious routing scheme for \emph{single-source} instances that achieve an $\Oh(\alpha \cdot \log^3 n / \log \log n)$ competitive ratio. In the process, we make several technical contributions which may be of independent interest. In particular, we give an efficient algorithm for computing \emph{low-radius decompositions} of directed graphs parameterized by balance. We also define and construct \emph{low-stretch arborescences}, a generalization of low-stretch spanning trees to directed graphs.
  
  On the negative side, we present new lower bounds for oblivious routing problems on directed graphs. We show that the competitive ratio of oblivious routing algorithms for directed graphs is $\Omega(n)$ in general; this result improves upon the long-standing best known lower bound of $\Omega(\sqrt{n})$ \cite{HajiaghayiKLR07}. We also show that our restriction to single-source instances is necessary by showing an $\Omega(\sqrt{n})$ lower bound for multiple-source oblivious routing in Eulerian graphs.

  We also study the maximum flow problem in balanced directed graphs with arbitrary capacities. We develop an efficient algorithm that finds an $(1+\epsilon)$-approximate maximum flows in $\alpha$-balanced graphs in time $\Oht(m \alpha^2 / \epsilon^2)$. We show that, using our approximate maximum flow algorithm, we can efficiently determine whether a given directed graph is $\alpha$-balanced. Additionally, we give an application to the directed sparsest cut problem.
}

%% file: intro.tex
\section{Introduction}
\label{sec:intro}

In this paper, we study several fundamental routing questions in \emph{directed graphs} that are \emph{nearly} Eulerian. We introduce  the notion of \emph{balance} for directed graphs that quantifies how far away a graph is from being Eulerian\footnote{A directed graph is Eulerian if, for each vertex, the total weight of its incoming edges is equal to the total weight of its outgoing edges. An equivalent definition is that for each cut $S \subseteq V$, the total weight of edges from $S$ to $V\setminus S$ is equal to the total weight of edges from $V\setminus S$ to $S$.}: a weighted directed graph is $\alpha$-balanced if for every cut $S \subseteq V$, the total weight of edges going from $S$ to $V\setminus S$ is within factor $\alpha$ of the total weight of edges going from $V\setminus S$ to $S$. Several important families of graphs are nearly balanced, in particular, Eulerian graphs (with $\alpha = 1$) and residual graphs of $(1+\epsilon)$-approximate undirected maximum flows (with $\alpha=\Oh(1/\epsilon)$).

We use the notion of balance to give a more fine-grained understanding of several well-studied routing questions that are considerably harder in directed graphs. The first question that we address is that of designing \emph{oblivious routing} schemes for directed graphs. Oblivious routing schemes were introduced in the seminal work of R{\"a}cke \cite{Racke02}. They are motivated by practical applications in routing traffic in massive networks such as the Internet, where it is necessary to route each request independently of the other requests and the current traffic in the network. Oblivious routing schemes were developed in a sequence of works \cite{Racke02,AzarCFKR03,BienkowskiKR03,HajiaghayiKLR05,HajiaghayiKLR06,HajiaghayiKLR07,Racke08,EnglertR09}. In particular, if the graph is undirected, there exist oblivious routing schemes that achieve competitive ratio $O(\log n)$ \cite{Racke08}, where $n$ is the number of nodes, and this result is optimal \cite{BartalL99,MaggsHVW97,MaggsHVW97}. In contrast, Hajiaghayi \etal \cite{HajiaghayiKLR07} show a strong lower bound of $\Omega(\sqrt{n})$ on the competitive ratio of routing obliviously in directed graphs. This lower bound holds even for \emph{single-source} instances of bounded degree graphs, as well as for instances with symmetric demands.

In this paper, we revisit oblivious routing in directed graphs, and we show that balanced graphs bridge the gap between directed and undirected graphs (see \Cref{sec:routing}). Our main algorithmic result is an oblivious routing scheme for \emph{single-source} instances that achieve an $\Oh(\alpha \cdot \log^3 n / \log \log n)$ competitive ratio. In the process, we make several technical contributions which may be of independent interest. In particular, we give an efficient algorithm for computing \emph{low-radius decompositions} of directed graphs parameterized by balance. We also define and construct \emph{low-stretch arborescences}, a new concept generalizing low-stretch spanning trees to directed graphs. Given the far-reaching implications of low-diameter decompositions and low-stretch spanning trees, we hope that our techniques may find other applications.

Our result is a generalization to directed graphs of R{\"a}cke's  influential work \cite{Racke08} that established a remarkable connection between oblivious routing in undirected graphs and metric embeddings into trees.

On the negative side, we present new lower bounds for oblivious routing problems on directed graphs. We show that the competitive ratio of oblivious routing algorithms for directed graphs has to be $\Omega(n)$ in general; this result improves upon the long-standing best known lower bound of $\Omega(\sqrt{n})$ \cite{HajiaghayiKLR07}. We also show that the restriction to single-source instances is necessary by showing an $\Omega(\sqrt{n})$ lower bound for multiple-source oblivious routing in Eulerian graphs.

The second question that we study is that of finding an \emph{approximate maximum flow} in balanced graphs. The maximum flow problem has received considerable attention in recent years, leading to several breakthrough results. This line of work has led to the development of almost linear time algorithms for approximate maximum flows in undirected graphs \cite{KelnerLOS14,Sherman13} and the subsequent improvement of \cite{Peng14,RackeST14}. In contrast, progress on directed graphs has been comparatively more modest, and the only improvements are the breakthrough results of Madry, yielding an $\Oht(m^{10/7})$-time algorithm for \emph{unit-capacity} directed graphs with $m$ edges \cite{Madry13} and of Lee and Sidford, obtaining a running time of $\Oht(m\sqrt{n})$ for arbitrary directed graphs \cite{LeeS13a}. These improve over the long-standing best running time of $\Oht(m \min(\sqrt{m}, n^{2/3}))$ given by Goldberg and Rao \cite{GoldbergR98}.

In this paper, we study the maximum flow problem in balanced directed graphs with arbitrary capacities (see \Cref{sec:maxflow}). We develop an efficient algorithm that finds an $(1+\epsilon)$-approximate maximum flows in $\alpha$-balanced graphs in time $\Oht(m \alpha^2 / \epsilon^2)$. Our algorithm builds on the work of Sherman \cite{Sherman13} and it can be viewed as an analogue of his result for directed graphs. The running time of our algorithm degrades gracefully with the imbalance of the graph and thus it suggests that balanced graphs provide a meaningful bridge between undirected and directed graphs.

We show that, using our approximate maximum flow algorithm, we can efficiently determine whether a given directed graph is $\alpha$-balanced (see \Cref{sec:imbalance}). Additionally, we give an application to the directed sparsest cut problem (see \Cref{sec:sparsest}).

\subsection{Related Work}

\noindent{\bf Oblivious Routing.}
Oblivious routing schemes are well-studied and several results are known; we refer the reader to \cite{Racke09} for a comprehensive survey of results for undirected graphs. As mentioned previously, in edge-weighted undirected graphs one can achieve a competitive ratio of $\Oh(\log{n})$ \cite{Racke08}, and it is the best possible \cite{BartalL99,MaggsHVW97,MaggsHVW97}. Hajiaghayi \etal \cite{HajiaghayiKLR07} studied oblivious routing schemes in node-weighted undirected graphs and directed graphs. Their work gives an $\Omega(\sqrt{n})$ lower bound on the competitive ratio for both node-capacitated undirected graphs and directed graphs. They also show that these lower bounds still hold in more restricted settings, such as single-source instances. On the positive side, they give oblivious routing scheme with competitive ratios of $\Oh(\sqrt{n} \log n)$ for single-source instances in bounded-degree directed graphs, and $\Oh(\sqrt{k} n^{1/4} \log{n})$ for general instances in directed graphs, where $k$ is the number of commodities and in the worst case $k = \Theta(n^2)$.

\noindent{\bf Maximum $s$-$t$ Flows.} The maximum flow problem is one of the most central problems in combinatorial optimization and has been studied extensively over the past several decades. Until recently, most approaches have been based on combinatorial methods such as augmenting paths, blocking flows, push-relabel, etc. This line of work culminated in the seminal algorithm of Goldberg and Rao \cite{GoldbergR98} that computes a maximum flow in time $O(\min(n^{2/3}, m^{1/2}) \log(n^2/m) \log{U})$ in directed graphs with integer weights that are at most $U$.

Over the past decade, a new approach emerged based on techniques drawn from several areas such as continuous optimization, numerical linear algebra, and spectral graph theory. These approaches led to a nearly-linear time algorithm for approximate maximum flows in \emph{undirected graphs} \cite{Sherman13,KelnerLOS14,Peng14}, an $\Oht(m^{10/7})$-time algorithm for maximum flows in \emph{unit-capacity directed graphs} \cite{Madry13} and an $\Oht(m\sqrt{n})$-time algorithm for arbitrary directed graphs \cite{LeeS13a}.

\subsection{Organization}

The rest of this paper is organized as follows. In Section~\ref{sec:overview}, we give an overview of our main results and introduce the definitions and notation we use throughout the paper. In Section~\ref{sec:routing}, we give our oblivious routing scheme for single-source instances. In Section~\ref{sec:lower}, we state our lower bounds for oblivious routing. In Section~\ref{sec:maxflow} we give our approximate maximum flow algorithm and applications. Many proofs are deferred to the Appendix.

%% file: overview.tex
\section{Overview}
\label{sec:overview}

\subsection{Basic Definitions}
%\noindent\textbf{Basic definitions.} 
We study directed graphs $G = (V, E, w, l)$ with edge set $E \subseteq V \times V$, edge weights $w: E \to \R_+$ and edge lengths $l: E \to \R_+$. Throughout this paper, we assume that $G$ is strongly connected. In several applications we deal with graphs without weights or lengths.
For graphs with edge lengths, we let $d(u, v)$ denote the shortest path distance from $u$ to $v$.

We associate the following matrices with the graph $G$. The matrix of edge weights is defined as $\mc\defeq\mdiag(w)$ and the vertex-edge incidence matrix $\mb\in\R^{V\times E}$ is defined as 
$
\mb_{s, (u,v)}\defeq
    -1$ if $s=u$, $1$ if $s=v$ and $0$ otherwise.
We are interested in finding flows that route demands with low congestion. The congestion incurred by a flow $f$ is $\normInf{\mc^{-1}f}$, and we say $f$ routes demands $b$ if $\mb f = b$.
The problem of finding a minimum congestion flow for a given demand vector, and its dual, the maximum congested cut, can be formulated as follows:

\begin{equation*}
    \begin{aligned}
        & \underset{f}{\text{min.}}
        & & \|\mc^{-1}f\|_\infty
        & \text{s.t.}
        & & \mb f = d,
          f \geq 0.\\
        & \underset{v}{\text{max.}}
        & & b^\top v
        & \text{s.t.}
        & & \|\mc\max(\mb^\top v, 0)\|_1 \leq 1.\\
    \end{aligned}
\end{equation*}

We let $OPT_b$ denote the optimum value of these problems.
Throughout the paper, we let $b_S = \sum_{u\in S} b_u$ and $w(S, T)$ denote the total weight of edges from $S$ to $T$.
It is well-known that for the second problem, one of the threshold cuts with respect to $v$ achieves $b_S/w(S, V - S)\geq b^\top v$.

\subsection{Balance}
 We parameterize strongly connected directed graphs by their \emph{imbalance}:

\begin{definition}[Imbalance]
\label{def:balanced}
    Let $G = (V, E, w)$ be a strongly connected directed graph.
    We define its \emph{imbalance}, $\bal(G)$, as the minimum $\alpha$ such that 
%    \begin{align*}
$
        w(S, V \setminus S) \leq \alpha \cdot w(V \setminus S, S)
$
%    \end{align*}
for every $S \subseteq V$.
\end{definition}

Two canonical families of balanced graphs are Eulerian graphs.
and residual graphs of approximate undirected maximum flows.

\begin{fact}
\label{fct:residual}
    A strongly connected directed graph $G$ is Eulerian if and only if $\bal(G) = 1$.
    If $G$ is the residual graph of a $(1+\epsilon)$-approximate undirected maximum flow, then $\bal(G) = \Oh(\epsilon^{-1})$.
\end{fact}

\begin{theorem}[Equivalent definitions of balance]
\label{thm:definitions}
Let $G = (V, E, w)$ be a directed graph.
The following statements are equivalent:
\begin{enumerate}
    \item $\bal(G) \leq \alpha$.
    \item There exists a circulation $f$ on $G$ with all edge congestions in $[1, \alpha]$.
    \item Let $d = \mb\vec1$ be the residual degrees in $G$.
        Then $-d$ can be routed with congestion $\alpha - 1$.
\end{enumerate}
\end{theorem}

\subsection{Oblivious Routing Schemes}
An oblivious routing scheme is a linear operator that, for each source-destination pair $(s, t) \in V \times V$, specifies how to route one unit of flow from $s$ to $t$ independently of the other pairs. Given a \emph{demand vector} $\vd: D \rightarrow \mathbb{R}_+$ on a set $D \subseteq V \times V$ of source-sink pairs, one can produce a multi-commodity flow that meets these demands by routing each demand pair using the (pre-specified) operator, independently of the other demands. The \emph{competitive ratio} of an oblivious routing scheme is the worst ratio among all possible demand vectors between the congestion of the multi-commodity flow given by the scheme and the congestion of the minimum congestion multi-commodity flow for the given demand vector.

\iffalse
Let $D \subseteq V \times V$ be a set of source-sink pairs.
An $\alpha$-competitive oblivious routing for $D$ is an assignment of a unit flow from $s$ to $t$ for each $(s, t) \in D$ such that for any weighted combination of $D$, the congestion of the sum of the assigned flows is within $\alpha$ of minimum possible.
\fi

Our main positive result concerning oblivious routings, given in \Cref{sec:routing}, is the existence of good single-source oblivious routings for balanced graphs.
A single-source oblivious routing with source $s \in V$ has $D = \{s\} \times V$.

\begin{theorem}[Single Source Oblivious Routings]
\label{thm:routing}
    Every strongly connected graph $G$ admits a single-source oblivious routing, from any source, with competitive ratio $\Oh(\bal(G) \cdot \log^3 n / \log \log n)$.
\end{theorem}

We achieve this result by generalizing an algorithm for undirected graphs given by Racke \cite{Racke08}.
The core difficulty that we need overcome is to find a good way to cluster the vertices of a directed balanced graph.
We define the \emph{radius} of a cluster $C \subseteq V$ as $\min_{u\in C}\max_{v\in C} d(u,v).$.
The \emph{volume} $\vol(G)$ of $G$ is defined as $\vol(G) \defeq \sum_{e \in E} l(e)w(e)$.
Our clustering algorithm is presented in \Cref{sec:ddecomp}, and its guarantees can be formalized as follows:

\begin{theorem}[Balanced Graph Clustering]
\label{thm:ddecomp}
    Let $G = (V, E, w, l)$ be a directed graph.
    Then for every $r > 0$, $V$ can be partitioned into clusters such that every cluster has radius at most $r$, and the total weight of edges going between different clusters is $\Oh(\bal(G)\vol(G)\log n / r)$.
    Moreover, such a partition can be found in expected linear time.
\end{theorem}

The guarantees of \Cref{thm:ddecomp} for undirected graphs match those given by prior work \cite{Awerbuch85, AlonKPW95, Bartal96, MillerPX13}.
Extending the statement to directed graphs is nontrivial, as it requires making the notion of cluster radii directed.

In \Cref{sec:lower} we give a new lower bound for all-pairs oblivious routings in directed graphs.

\begin{restatable}{theorem}{lowerbound}
\label{thm:lowerbound}
    No oblivious routing algorithm for directed graphs can guarantee competitive ratio better than $\Omega(n)$.
\end{restatable}

We also show that restricting ourselves to single-source oblivious routings is necessary to achieve a small competitive ratio even when $\bal(G)=1$.

\begin{restatable}{theorem}{lowerboundbalanced}
\label{thm:lowerboundbalanced}
    No oblivious routing algorithm for Eulerian graphs can guarantee competitive ratio better than $\Omega(\sqrt{n})$.
\end{restatable}

%\smallskip
%\noindent\textbf{Maximum flows.} 
\subsection{Maximum Flows}
Finally, we consider the maximum $s$-$t$ flow problem in directed graphs parameterized by balance. Given a source $s$ and a destination $t$, the maximum $s$-$t$ flow problem asks us to find a flow $f$ that routes as much flow as possible from $s$ to $t$ while sending at most $w_e$ units of flow along each edge $e$. In \Cref{sec:maxflow} we show the following result.

\begin{theorem} [Approximate Maximum Flow]
\label{thm:maxflow}
    Given a strongly connected directed graph $G$, a source $s$, and a sink $t$ there is an algorithm that finds a $(1+\epsilon)$-approximate maximum $s$-$t$ flow and a $(1-\epsilon)$-approximate minimum $s$-$t$ cut in $G$ in time $\Oht(m\cdot \bal(G)^2 / \epsilon^{2})$.
\end{theorem}

To achieve quadratic dependency on $\epsilon$, in \Cref{sec:proximal} we provide a general analysis of gradient descent for composite function minimization under non-Euclidean norms.

We also show applications of this result to computing the sparsest cut (\Cref{sec:sparsest}) and we prove the following result on computing the imbalance of a graph (\Cref{sec:imbalance}).

\begin{restatable}{lemma}{isbalanced}
\label{lem:isbalanced}
    There is an algorithm that either certifies that $\bal(G) \leq \alpha$ or shows that $\bal(G) > (1-\epsilon)\alpha$ in time $\Oht(m \alpha^2 / \epsilon^{2})$.
\end{restatable}

%% file: routing.tex
\section{Oblivious Routing on Balanced Graphs}
\label{sec:routing}

\input{ddecomp}
\input{lowstretch}
\input{racke}

%% file: ddecomp.tex
\subsection{Low-radius Decompositions}
\label{sec:ddecomp}

Our algorithm for clustering directed graphs, presented in \Cref{fig:ddecomp}, is based on the scheme given by Miller, Peng and Xu \cite{MillerPX13}.
We first pick a start time $x_v$ for every vertex $v$ from an exponential distribution, and then explore the graph, starting the search from $v$ at time $x_v$ and proceeding at unit speed.
Each vertex $u$ is assigned to the vertex $v$ that reached it first.

\begin{figure}[ht]
	\noindent
	\centering
	\fbox{
		\begin{minipage}{6in}
			\noindent $(V_1, V_2, \ldots) = \textsc{Cluster-Directed} (G, r)$, where $G = (V, E, l)$ is a directed graph and $r > 0$.
			\begin{enumerate}
				\item Set $\beta := \log n / (10r)$.
                \item For every vertex $v \in V$ pick $x_v \sim \rm{Exp}(\beta)$.\footnotemark
				\item For each vertex $u \in V$, assign $u$ to the cluster rooted at the vertex $v \in V$ which minimizes $-x_v + d(v, u)$.
				\item If any of the clusters has radius greater than $r$, return to step $2$. Otherwise, return the clusters.
			\end{enumerate}
		\end{minipage}
	}
	\caption{The low-radius decomposition algorithm for directed graphs.}
	\label{fig:ddecomp}
\end{figure}

\footnotetext{$\rm{Exp}(\beta)$ is the exponential distribution with parameter $\beta$, with p.d.f. $f(x) = \beta e^{-\beta x}$ on $x \geq 0$.}

\begin{figure}[ht]
\noindent
\centering
\input{fig1.tex}
\caption{An unweighted Eulerian graph where a particular edge is very likely to be cut by the scheme of \cite{MillerPX13}.}
\label{fig:mpxexample}
\end{figure}
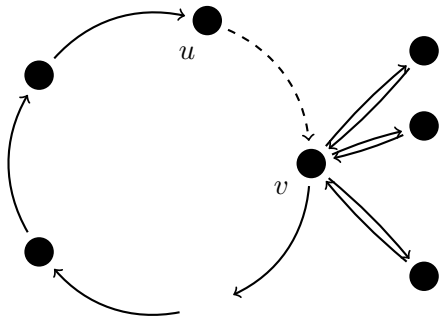

Our goal is to show that this procedure \emph{cuts} few edges, i.e. assigns the endpoints of few edges to different clusters. The original analysis of \cite{MillerPX13} shows that for undirected graphs, this approach guarantees cutting each edge $e$ with low probability, namely $\Oh(l(e) \log n / r)$.
It turns out that even in the case of unweighted Eulerian graphs such a guarantee no longer holds; there may exist edges that are cut with very high probability.
Consider for instance (\Cref{fig:mpxexample}) a directed cycle of length $3^{k}$, with an undirected star of $2^{k^2}$ leaves attached to one of its vertices, $v$.
Set $r := 2^k$.
Let $u$ be the vertex preceding $v$ on the cycle.
It is now easy to verify by calculation that the edge $(u, v)$ is cut with probability arbitrarily close to $1$ for a large enough $k$.
With high probability, $v$ will be contained in a cluster rooted at one of the $2^{k^2}$ leaves attached to it; also with high probability, no such cluster will contain $u$.

This issue requires us to find a new way to guarantee that the total weight of cut edges is low.
Our key idea is to show that, for any fixed \emph{cycle}, the expected number of edges in the cycle that are cut is small.
The desired guarantees then follow by noting that any graph $G$ can be approximated up to a factor $\bal(G)$ by a sum of cycles (\Cref{thm:definitions}).

\begin{lemma}
    \label{lem:cycle}
    Let $\mathcal{P}$ be the partition returned by $\textsc{Cluster-Directed}(G, r)$.
    For any simple cycle $C$ in $G$, the expected number of edges in $C$ that go between different clusters in $\mathcal{P}$ is an $\Oh(\log n / r)$ fraction of the length of $C$.
\end{lemma}

As the above example demonstrates, we cannot base the proof of \Cref{lem:cycle} on the \emph{location} of the cuts, as it might depend strongly on the input graph.
However, we can prove that, intuitively, cuts occur \emph{infrequently} as the graph is explored.
This is the crucial idea of the proof: we analyze the occurrence of cuts over time rather than bounding the probabilities of particular cuts.
Then we use the fact that a cycle of length $L$ is fully explored within $L$ time steps after the time it is visited for the first time.
The analysis is presented in Appendix~\ref{sec:ddecomp_proofs}.

%% file: fig1.tex
\begin{tikzpicture}[thick,
  every node/.style={draw,circle},
  snode/.style={fill=black},
  invnode/.style={inner sep=0pt,opacity=0,text opacity=1},
  ->,shorten >= 3pt,shorten <= 3pt
]

\def \n {5}

\foreach \i in {3,4}
{
    \node[snode] (s\i) at ({360/\n * (\i - 1)}:2) {};
}
\node[snode] (s2) at ({360/\n * (2 - 1)}:2) [label={260: $u$}] {};
\node[snode] (s1) at ({360/\n * (1 - 1)}:2) [label={210: $v$}] {};
\node[invnode] (s5) at ({360/\n * (5 - 1)}:2) {\ldots};

\node[snode] (t1) at (3.5,1.5) [] {};
\node[snode] (t2) at (3.5,0.5) [] {};
\node[invnode] (t3) at (3.5,-0.5) [] {\vdots};
\node[snode] (t4) at (3.5,-1.5) [] {};

\draw[] (s5) edge[bend left] (s4);
\draw[] (s4) edge[bend left] (s3);
\draw[] (s3) edge[bend left] (s2);
\draw[] (s2) edge[dashed,bend left] (s1);
\draw[] (s1) edge[bend left] (s5);

\draw[] (s1) edge[bend left=5] (t1);
\draw[] (t1) edge[bend left=5] (s1);
\draw[] (s1) edge[bend left=5] (t2);
\draw[] (t2) edge[bend left=5] (s1);
\draw[] (s1) edge[bend left=5] (t4);
\draw[] (t4) edge[bend left=5] (s1);

\end{tikzpicture}

%% file: lowstretch.tex
\subsection{Low-stretch Arborescences}
\label{sec:lowstretch}

Let $G$ be a directed graph and let $s$ be a vertex in $G$. We say that a directed graph $T$ is an \emph{arborescence} rooted at $s$ for every vertex $v$, there is a unique directed path in $T$ from $s$ to $v$. In this section, we define and construct \emph{low-stretch arborescences}, which are a key intermediate step between low-radius decompositions and oblivious routings.

\begin{definition}
Let $G = (V, E, w, l)$ be a directed graph.
We define the \emph{stretch} of an edge $(u, v) \in E$ with respect to an arborescence $T$ on the vertex set $V$ as
$
    w(u, v) \cdot d_T(u, v),
$
where $d_T(u, v)$ is the distance between $u$ and $v$ in the \textbf{undirected} tree corresponding to $T$.
\end{definition}

Following the notation of \cite{Racke08}, we define the \emph{load}, $\mbox{load}_T(e)$, of an edge $e \in T$ as the sum of the weights of edges $(u, v) \in E(G)$ such that $e$ is on the path between $u$ and $v$ in the \emph{undirected} tree corresponding to $T$. Note that the total load of the edges in $T$ is equal to the total stretch of the edges in $G$.

In order to construct low-stretch arborescences, we will recursively cluster $V$ using the algorithm from the previous section.
The algorithm $\textsc{Find-Arborescence}$ is defined and analyzed in Appendix~\ref{sec:lowstretch_proofs}.
It is similar to the scheme given by Bartal \cite{Bartal96}.
One major difficulty is that the clusters returned by $\textsc{Cluster-Directed}$ may be very imbalanced; in particular, they need not be strongly connected.
In order to resolve this issue, we introduce the notion of \emph{additive imbalance} and prove that our clustering algorithms still give good guarantees for graphs with low additive imbalance.

\begin{theorem}
\label{thm:lowstretch}
    Let $G = (V, E, w, l)$ be a strongly connected directed graph.
    Let $s \in V$.
    Let $T = \textsc{Find-Arborescence}(G, s)$.
    Then:
    \begin{itemize}
        \item $T$ has vertex set $V$ and is rooted at $s$,
        \item every arc $(u, v)$ in $T$ can be mapped to a path from $u$ to $v$ in $G$ of equal length, and
        \item the expected total stretch of $G$ with respect to $T$ is $\Oh(\bal(G) \vol(G) \log^3 n / \log \log n)$.
    \end{itemize}
    Moreover, the algorithm works in expected $\Oh(m \log n)$ time.
\end{theorem}

%% file: racke.tex
\subsection{Constructing the Routing}
\label{sec:racke}

\begin{figure}[ht]
\noindent
\centering
\fbox{
\begin{minipage}{6in}
    \noindent $((T_1, \lambda_1), \ldots, (T_k, \lambda_k)) = \textsc{Find-Routing}(G, s)$ where $G = (V, E, w)$ is a strongly connected directed graph and $s \in V$.
\begin{enumerate}
    \item Set $k := 0$ and $p^{(0)}_e := 1$ for all $e \in E$.
    \item While $\sum_{i=1}^k \bv{\lambda}_i < 1$:
        \begin{enumerate}
            \item $k := k + 1$.
            \item Let $G_k = (V, E, l_k)$ be a copy $G$ with edge lengths
                \begin{align*}
                    l_k(e) := p^{(k-1)}_e / \left(w(e) \sum_{e'} p^{(k-1)}_{e'}\right).
                \end{align*}
            \item $T_k := \textsc{Find-Arborescence}(G, s)$ (pick the minimum-stretch arborescence out of $\Oh(\log n)$ runs).
            \item $\ell_k := \max_e\{\mbox{load}_{T_k}(e) / w(e)\}$.
            \item $\lambda_k := \min\left(1/\ell_k, 1 - \sum_{i=1}^{k-1} \lambda_i\right)$.
            \item For all edges $e$ set:
                \begin{align*}
                    p^{(k)}_e := p^{(k-1)}_e \cdot \exp(\lambda_k \cdot \mbox{load}_{T_k}(e) / w(e)).
                \end{align*}
        \end{enumerate}
    \item Return $((T_1, \lambda_1), \ldots, (T_k, \lambda_k))$.
\end{enumerate}
\end{minipage}
}
\caption{The algorithm for finding single-source oblivious routings on balanced graphs (adapted from \cite{Racke08}).}
\label{fig:racke}
\end{figure}

Given an algorithm for constructing low-stretch arborescences, we can use it to compute a good oblivious routings using the approach proposed by \cite{Racke08}. The oblivious routing that we construct for a given source $s$ will be a convex combination of arborescences rooted at $s$, with the flow for demand $(s, u)$ being defined as the convex combination of the corresponding paths. The algorithm is given in \Cref{fig:racke}. 

The key idea we employ to extend the analysis of the algorithm to a  directed graph $G$ is to prove that the routing scheme we construct is competitive even for the \emph{undirected} graph underlying $G$.

\begin{lemma}[\cite{Racke08}, adapted]
\label{lem:racke}
    Let $G$ be a strongly connected directed graph and $s$ be a vertex in $G$.
    Let $((T_1, \lambda_1), \ldots, (T_k, \lambda_k)) := \textsc{Find-Routing}(G, s)$.
    Then with high probability $((T_1', \lambda_1), \ldots, (T_k',  \lambda_k))$ is an $\Oh(\bal(G) \log^3 n / \log \log n)$-competitive oblivious routing for $G'$, where $T_1', \ldots, T_k', G'$ are the undirected counterparts of $T_1, \ldots, T_k$ and $G$, respectively, that we obtain by ignoring the directions.
\end{lemma}

In order to finish the analysis, we only need to note that $((T_1, \lambda_1), \ldots, (T_k, \lambda_k))$ is an oblivious routing for $G$.

\Proofof{\Cref{thm:routing}}
We prove that for any $s$, the output of $\textsc{Find-Routing}(G, s)$ satisfies the criteria stated in the theorem statement.
It follows from \Cref{lem:racke} that with high probability, $((T_1', \lambda_1), \ldots, (T_k', \lambda_k))$ is an $\Oh(\bal(G) \log^3 n / \log \log n)$-competitive oblivious routing for $G'$, where $T_1', \ldots, T_k', G'$ are undirected counterparts of $T_1, \ldots, T_k$, $G$, respectively.
In particular, it is also an $\Oh(\bal(G) \log^3 n / \log \log n)$-competitive oblivious routing from $s$.
    Now it is enough to observe that since $T_1, \ldots, T_k$ are directed away from $s$, $((T_1, \lambda_1), \ldots, (T_k, \lambda_k))$ is an oblivious routing from $s$ in $G$.
    Since it is $\Oh(\bal(G) \log^3 n / \log \log n)$-competitive in $G'$, it must also be $\Oh(\bal(G) \log^3 n / \log \log n)$-competitive in $G$.
\QED

%% file: lower.tex
\section{Lower Bounds}
\label{sec:lower}

We prove new lower bounds for oblivious routings in directed graphs.
The constructions and proofs are given in Appendix~\ref{sec:lower_proofs}.

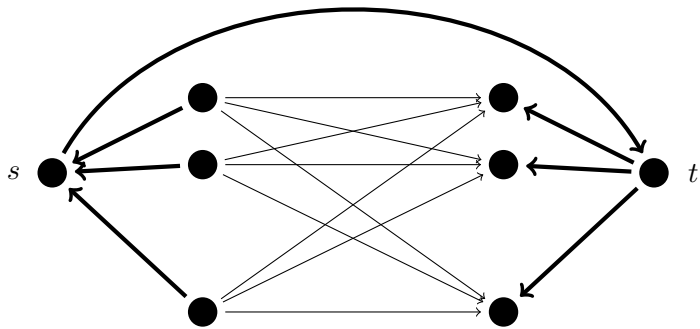
\begin{figure}[ht]
\noindent
\centering
\input{fig2.tex}
\caption{The example from \Cref{thm:lowerbound}. The thick edges have weight $n$, the other edges have weight $1$. Any oblivious routing must put too much flow on the edge $(s, t)$ when routing between the vertices of the biclique.}
\label{fig:lowerbound}
\end{figure}
\lowerbound*

\begin{figure}[ht]
\noindent
\centering
\input{fig3.tex}
\caption{The example from \Cref{thm:lowerboundbalanced}. The thick edges have weight $\sqrt{n}$, the other edges have weight $1$. Any oblivious routing must put too much flow on the outer cycle when routing between consecutive vertices of the inner cycle.}
\label{fig:lowerboundbalanced}
\end{figure}
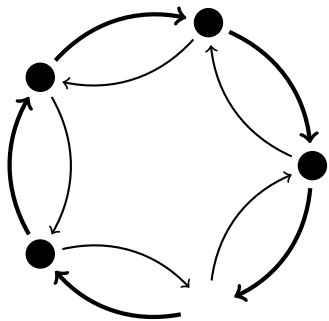
\lowerboundbalanced*

%% file: fig2.tex
\begin{tikzpicture}[thick,
  every node/.style={draw,circle},
  snode/.style={fill=black},
  invnode/.style={inner sep=0pt,opacity=0,text opacity=1},
  ->,shorten >= 3pt,shorten <= 3pt
]

\node[snode, xshift = -2cm, yshift=-1cm] (s) [label=left: $s$] {};

% the vertices of S
\begin{scope}[start chain=going below,node distance=5mm]
\foreach \i in {1,2}
    \node[snode,on chain] (s\i) [] {};
\node[invnode,on chain] (sdots) [] {\vdots};
\foreach \i in {3}
    \node[snode,on chain] (s\i) [] {};
\end{scope}

\node[snode, xshift = 6cm, yshift=-1cm] (t) [label=right: $t$] {};

% the vertices of T
\begin{scope}[xshift=4cm,start chain=going below,node distance=5mm]
\foreach \i in {1,2}
    \node[snode,on chain] (t\i) [] {};
\node[invnode,on chain] (tdots) [] {\vdots};
\foreach \i in {3}
    \node[snode,on chain] (t\i) [] {};
\end{scope}

% the edges
\foreach \i in {1,...,3}
    \foreach \j in {1,..., 3}
        \draw (s\i) edge[thin] (t\j);

\foreach \i in {1,...,3}
\draw (s\i) edge[ultra thick] (s);

\foreach \i in {1,...,3}
    \draw (t) edge[ultra thick] (t\i);

\draw (s) edge[ultra thick,bend left=60] node[invnode] [] {} (t);
\end{tikzpicture}

%% file: fig3.tex
\begin{tikzpicture}[thick,
  every node/.style={draw,circle},
  snode/.style={fill=black},
  invnode/.style={inner sep=0pt,opacity=0,text opacity=1},
  ->,shorten >= 3pt,shorten <= 3pt
]

\def \n {5}

\foreach \i in {1,...,4}
{
    \node[snode] (s\i) at ({360/\n * (\i - 1)}:2) {};
}
\node[invnode] (s5) at ({360/\n * (5 - 1)}:2) {\ldots};

\draw[] (s1) edge[bend left] (s2);
\draw[] (s2) edge[bend left] (s3);
\draw[] (s3) edge[bend left] (s4);
\draw[] (s4) edge[bend left] (s5);
\draw[] (s5) edge[bend left] (s1);

\draw[] (s5) edge[ultra thick, bend left] (s4);
\draw[] (s4) edge[ultra thick, bend left] (s3);
\draw[] (s3) edge[ultra thick, bend left] (s2);
\draw[] (s2) edge[ultra thick, bend left] (s1);
\draw[] (s1) edge[ultra thick, bend left] (s5);

\end{tikzpicture}

%% file: maxflow.tex
\section{Maximum Flow and Applications}
\label{sec:maxflow}

\subsection{Directed Maximum Flow}

In this subsection we show how to efficiently compute an $(1+\epsilon)$-approximate maximum flow in directed graphs given a good \emph{congestion-approximator}.

\begin{definition}
    An $\alpha$-congestion-approximator for $G$ is a matrix $\mr$ such that for any demand vector $b$,
    $
        \normInf{\mr b} \leq OPT_b \leq \alpha \normInf{\mr b}.
    $
\end{definition}

Since $\normInf{\mr b} = \normInf{-\mr b}$, only well-balanced graphs admit good congestion approximators:
\begin{fact}
\label{fac:congbalance}
    If $G$ admits an $\alpha$-congestion approximator, $\bal(G) \leq \alpha$.
\end{fact}

For undirected graphs, $\Oht(1)$-congestion-approximators can be computed in nearly linear time \cite{Madry10,Sherman13,KelnerLOS14,Peng14}. This implies that for directed $G$ we can compute $\Oht(\bal(G))$-congestion-approximators in nearly linear time by the following fact:

\begin{fact}
    Let $G$ be a directed graph and $G'$ be its undirected copy.
    Then for any demand vector $b$
    $
        OPT_b(G') \leq OPT_b(G) \leq (1 + \bal(G))OPT_b(G').
    $
\end{fact}

Our main result is the following:

\begin{theorem}
    Let $G$ be a directed graph.
    Given an $\alpha$-congestion-approximator $\mr$, we can compute an $(1+\epsilon)$-approximate maximum flow and minimum congested cut for any demand vector in time $\Oht(m \alpha^2 / \epsilon^{2})$, assuming multiplication by $\mr$ and $\mr^\top$ can be done in $\Oht(m)$ time.
\end{theorem}

Our algorithm is based very heavily on the approach for undirected graphs given by Sherman \cite{Sherman13}. The main difference is the implementation of the key optimization procedure, presented in \Cref{fig:maxflow}. Due to space constraints, in this section we only outline the main changes needed to extend the algorithm of \cite{Sherman13} to balanced graphs.

Let $G$ be a directed graph and $b$ be a demand vector.
Assume we are given an $\alpha$-congestion-approximator $\mr$. Let $\lmax(x) \defeq \ln \sum_i (e^{x_i} + e^{-x_i})$ and define
\begin{align*}
    \mu(f)&\defeq\lmax(2\alpha\mr(b - \mb f))\\
    \phi(f)&\defeq \normInf{\mc^{-1}f} + \mu(f)
\end{align*}

\begin{figure}[ht]
\noindent
\centering
\fbox{
\begin{minipage}{6in}
    \noindent $(f,v) = \textsc{Almost-Route-Directed} (b, \epsilon, f_0)$
\begin{enumerate}
    \item Initialize $f := f_0, \delta := \frac{\epsilon}{10\alpha^2}$.
    \item Scale $f$ and $b$ so that $\normInf{\mc^{-1} f} + 2\alpha\normInf{\mr(b - \mb f)} = 20\epsilon^{-1}\ln n$.
\item Repeat while any of the following conditions is satisfied:
    \begin{enumerate}
        \item if $\phi(f) < 16\epsilon^{-1} \ln n$, scale $f$ and $b$ up by $17/16$ and restart step $3$.
        \item let $s$ be $w(e)$ on the coordinates $e$ where $\nabla \mu(f)$ is negative and $0$ elsewhere. If $- \nabla \mu(f)^\top s > 1 + \frac{\epsilon}{4}$, set $f := f + \delta s$ and restart step $3$.
        \item if $\normInf{\mc^{-1} f} + \nabla \mu(f)^\top f > \frac{\epsilon}{4}$, set $f := f - \delta f$ and restart step $3$.
    \end{enumerate}
\item Set $x := 2\alpha\mr(b-\mb f)$.
\item Set $p := \nabla \lmax(x)$.
\item Set $v := \mr^\top p$.
\end{enumerate}
\end{minipage}
}
\caption{The algorithm for computing the maximum flow and minimum congested cut.}
\label{fig:maxflow}
\end{figure}

\begin{lemma}
\label{lem:maxflow-correct}
    After $\textsc{Almost-Route-Directed}(b, \epsilon,f_0)$ terminates, we have
    \begin{align*}
        \phi(f) \leq (1+\epsilon)\frac{b^\top v}{\|\mc\max(\mb^\top v, 0)\|_1},
    \end{align*}
    assuming $\epsilon \leq 1/2$.
\end{lemma}

\begin{lemma}
\label{lem:maxflow-runtime}
$\textsc{Almost-Route-Directed}(b, \epsilon,f_0)$ terminates within $\Oht(\log(1 + \epsilon_0) \alpha^2 / \epsilon^{3})$ iterations, where $\epsilon_0 = \max(\phi(f_0) / OPT_b - 1, \epsilon)$, assuming $\epsilon \leq 1/2$.
\end{lemma}

Note that \Cref{lem:maxflow-correct} implies that $v$ is a potential vector for a $(1+\epsilon)$-approximate minimum congested cut.
In order to recover the corresponding flow, we can employ the recursion described in \cite{Sherman13}.
The only additional component necessary for directed graphs is an $\Oh(\poly{n, \alpha})$-competitive oblivious routing.
Since by \Cref{fac:congbalance} it must be that $\alpha \geq \bal(G)$, this can be obtained easily by taking the maximum spanning in- and out-arborescences from any fixed vertex.

\begin{figure}[ht]
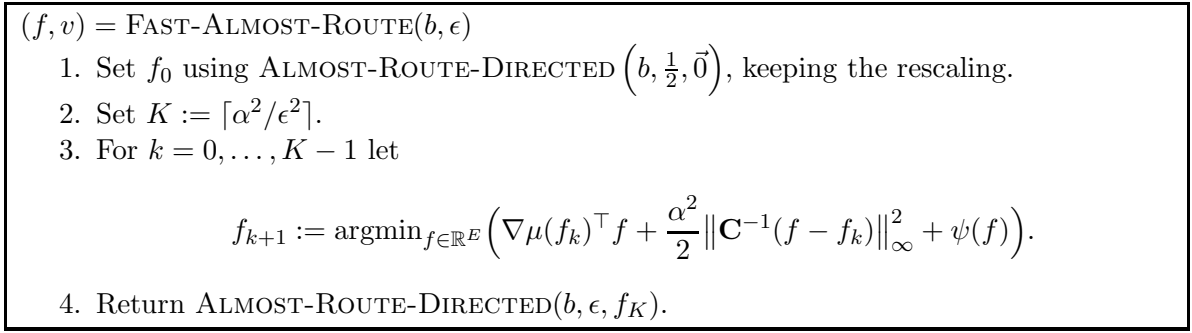

\noindent
\centering
\fbox{
\begin{minipage}{6in}
    \noindent $(f,v) = \textsc{Fast-Almost-Route} (b, \epsilon)$
\begin{enumerate}
    \item Set $f_0$ using $\textsc{Almost-Route-Directed}\left(b, \frac{1}{2}, \vec0\right)$, keeping the rescaling.
    \item Set $K := \lceil \alpha^2 / \epsilon^{2} \rceil$.
    \item For $k = 0, \ldots, K - 1$ let
        \begin{align*}
            f_{k+1} :=  \argmin_{f \in \R^E} \Big( \nabla \mu(f_k)^\top f + \frac{\alpha^2}{2} \normInf{\mc^{-1}(f - f_k)}^2 + \psi(f) \Big).
        \end{align*}
    \item Return $\textsc{Almost-Route-Directed}(b, \epsilon, f_K)$.
\end{enumerate}
\end{minipage}
}
\caption{Faster algorithm for computing the maximum flow and minimum congested cut.}
\label{fig:maxflow-fast}
\end{figure}

If we run $\textsc{Almost-Route-Directed}$ with $f_0 = \vec0$, we can find $(1 + \epsilon)$-approximate solutions in time $\Oht(m \alpha^2 / \epsilon^{3})$.
In order to improve the dependency on $\epsilon$, we can employ a general form of composite function minimization, introduced in \Cref{sec:proximal}.
Define
\begin{align*}
    \psi(f)&\defeq\begin{cases}
        \infty & \text{if for some } $e$, f_e / w(e)  \notin [0, 50 \ln n / \epsilon]\\
        \normInf{\mc^{-1} f} & \text{otherwise.}
\end{cases}
\end{align*}
The faster algorithm is presented in \Cref{fig:maxflow-fast}.

If we apply the analysis from Section~\ref{sec:proximal} (encapsulated in Theorem~\ref{thm:grad-desc-progress}), we obtain the following.

\begin{lemma}
    \label{lem:fastroute}
    $\textsc{Fast-Almost-Route}(b, \epsilon)$ terminates in $\Oht(m\alpha^2 / \epsilon^{2})$ time, assuming $\epsilon \leq 1/2$.
\end{lemma}

\subsection{Computing Imbalance}
\label{sec:imbalance}

As verifying balance can be reduced to a maximum flow computation by \Cref{thm:definitions}, we obtain the following result:

\isbalanced*

\subsection{Application to Directed Sparsest Cut}
\label{sec:sparsest}

In this subsection, assume $G = (V, E)$ is a directed graph that is unweighted, strongly connected, simple, and with an even number of vertices. We define the sparsity of a cut $(S, V \setminus S)$ as
$
    \frac{w(S, V\setminus S)}{|S|\cdot|V\setminus S|},
$
where $w(S, V \setminus S)$ is the number of edges going from $S$ to $V \setminus S$.
Note that under this definition, no cut can have sparsity greater than one.

\iffalse
We note that graphs that have no sparse cuts are necessarily well-balanced.
This lets us obtain the following result:
\fi

As a second application of our maximum flow algorithm, we get the following sparsest cut algorithm. While blocking flows could also possibly be used for our purpose, our approach is clean and may easily generalize to weighted graphs. We defer the details to the appendix.

\begin{lemma}
\label{lem:sparsest}
    Given $\phi \leq 1$, we can find a cut of sparsity $\phi$ in $G$ or determine that all cuts in $G$ have sparsity $\Omega(\phi / \log^2 n)$ in time $\Oht(m / \phi^{2})$.
\end{lemma}

%% file: proximal.tex
\subsection{Composite Function Minimization}
\label{sec:proximal}
In this section, we provide a non-Euclidean gradient descent method for minimizing a composite function $f(x) \overset{\mathrm{def}}{=} g(x) + \psi(x)$, where $g$ and $\psi$ have specific properties. The algorithm and its convergence guarantee are encapsulated in the following theorem, and they build on several works in convex optimization, such as \cite{Nesterov13,RichtarikT14}.

\iffalse
Here we provide a non-euclidean version of composite function minimization,
what I informally (and somewhat incorrectly) refer to as proximal
gradient. This was inspired by lecture notes and some neat proof tricks
in a Nesterov paper on composite function minimization and some insights
from some papers of Richtarik and more.\footnote{We should cite this formally.}
\footnote{Note, that when I wrote this note I was thinking of multiple projects
and also sent it to another group for a different project. As such
some of hte writing might be a little old and reflective of that notaiton
and perspective.}The main result of this section is the following:
\fi

\begin{theorem}
%[Non-Euclidean Gradient Descent for Composite Function Minimization ]
\label{thm:grad-desc-progress}
  Let $f\,:\,\R^{n}\rightarrow\R$ be a convex function given by $f(x) \overset{\mathrm{def}}{=} g(x)+\psi(x)$ where $g$ is convex and $L$-smooth\footnote{A function is $L$-smooth with respect to the norm $\| \cdot \|$ if, for all $\vx$ and $\vy$, $\| \nabla f(\vx) - \nabla f(\vy) \| \leq L \| \vx - \vy\|$. } with respect to some norm $\norm{\cdot}$. Moreover, assume that $f(x)$ is only finite on some region of diameter $D$ in $\norm{\cdot}$. Starting with some $x_{0}\in\R^{n}$ for all $k$ let 
\[
x_{k+1}:=\argmin_{x\in\R^{n}} \Big(\innerProduct{\grad g(x_{k})}{x}+\frac{L}{2}\norm{x- x_{k}}^{2}+\psi(x)\Big) \,.
\]
Then for all $k\geq1$ we have
\[
\epsilon_{k}\leq\max\left\{ \frac{2\cdot L\cdot D^{2}}{\lfloor\frac{k-1}{2}\rfloor+4}\,,\,\left(\frac{1}{2}\right)^{\lfloor\frac{k-1}{2}\rfloor}\epsilon_{0}\right\} 
\]
where $\epsilon_{k}=f(x_{k})-\min_{x}f(x)$.
%\footnote{Note that there is possibly a cleaner way to deal with the $\lfloor\frac{k-1}{2}\rfloor$ that appears in both but this Theorem would suffice for many purposes.} 
\end{theorem}
Note that the norm we use is arbitrary and we get a gradient
descent analysis without appealing to the dual norm. Also 
we do not require convex $\psi$ we only require convex $f$.

%% file: overview_proofs.tex
\section{Missing proofs from \Cref{sec:overview}}
\label{sec:overview_proofs}

\begin{lemma}
\label{lem:flowsymmetry}
    Let $G$ be a strongly connected directed graph.
    If demand vector $d$ can be routed in $G$ with congestion $c$, then $-d$ can be routed in $G$ with congestion at most $\bal(G)\cdot c$.
\end{lemma}
\Proof

Note that for any $v \in \mathbb{R}^n$
\begin{align*}
    \|{\bv U}\max({\bv B}v, 0)\|_1 \leq \bal(G)\|{\bv U}\max(-{\bv B}v, 0)\|_1
\end{align*}
follows from the definition of balance.

Hence it is easily seen that the optimum value for the dual problem is within a factor $\bal(G)$ for demands $d$ and $-d$.
Our theorem now follows from strong duality to the original problem.
\QED

\begin{lemma}
\label{lem:convexsets}
    Let $l, r \in \mathbb{R}$ with $l \leq r$.
    Let $C \subseteq \mathbb{R}^m$ be a convex set such that for any $S \subseteq \{1,2,\ldots,m\}$ there exists a point $x \in C$ such that $x_i$ is at least $l$ for $i \in S$ and $x_i$ is at most $r$ for $i \notin S$.
    Then there exists a point in $C$ with all coordinates in $[l, r]$.
\end{lemma}
\Proof
Let $P_i(S)$, for $i \in \{0, \ldots, m\}, S \subseteq \{1, \ldots, m\}$ be the subset of points $x \in C$ that satisfy
\begin{itemize}
    \item $x_j \in [l, r]$ for $j \leq i$ and
    \item $x_j \geq l$ for $j \in S$ and
    \item $x_j \leq r$ for $j \notin S$.
\end{itemize}

We prove that $P_i(S)$ is nonempty for every $i$ and $S$ by induction on $i$.
The base case $i = 0$ follows from the assumption on $C$.
Assume $i \in \{0, \ldots, m - 1\}$ and the thesis holds for $i$.
Let $S$ be any subset of $\{1, \ldots, m\}$.
Let $S_L := S \cup \{i + 1\}, S_R = S \setminus \{i + 1\}$.
Pick any $x_L \in P_i(S_L)$ and $x_R \in P_i(S_R)$.
Then a convex combination of $x_L$ and $x_R$ must belong to $P_{i+1}(S)$.
Since $S$ was arbitrary, this concludes the proof.
\QED

\Proofof{\Cref{thm:definitions}}
The implication $(2. \to 1.)$ and the equivalence $(2. \leftrightarrow 3.)$ are easy to check (note that the circulation of $2.$ is the sum of $\vec1$ and a routing of $-d$.). We now prove that if $\bal(G) \leq \alpha$ there exists a circulation in $G$ with each congestion in $[1, \alpha]$.

Note that for any subset $S$ of edges of $G$ we can route the residual degree $d_S$ induced by these edges with congestion $1$.
Hence by \Cref{lem:flowsymmetry} we can route $-d_S$ with congestion at most $\alpha$.
Adding these flows yields a circulation with congestion in $[1, \alpha + 1]$ on edges in $S$ and in $[0, \alpha]$ on the other edges.
Since the choice of $S$ was arbitrary, the thesis follows by \Cref{lem:convexsets}.
\QED

We now prove the following lemma, implying \Cref{fct:residual}.
\begin{lemma}
\label{lem:residual}
    Let $G = (V, E, w)$ be an undirected graph and $s, t \in V$.
    Let $d$ be a demand vector that can be routed in $G$ with congestion at most $1$.
    Let $f$ be a flow from $s$ to $t$ in $G$ with congestion not exceeding $1$ satisfying demands $(1-\epsilon)d$. 
    Let $H$ be the residual graph of $f$ in $G$.
    Then $\bal(H) \leq \left(2\epsilon^{-1}-1\right)$.
\end{lemma}
\Proof
We use the third equivalent definition of balance from \Cref{thm:definitions}.
The residual degrees in $H$ are $2(\epsilon-1)d$.
Since there exists a flow satisfying demands $\epsilon d$ with congestion $1$, $2(1-\epsilon)d$ can be routed in $H$ with congestion $2(1-\epsilon)\epsilon^{-1} = 2\epsilon^{-1} - 2$.
\QED

%% file: ddecomp_proofs.tex
\section{Missing proofs from \Cref{sec:ddecomp}}
\label{sec:ddecomp_proofs}

Before we prove \Cref{lem:cycle}, we shall study the properties of a class of two-way infinite sequences.

For better understanding, we now attempt to provide the intuition on how the sequences defined below are used in the proof.
For simplicity, assume we are trying to analyze the clustering of a \emph{path} rather than a cycle.
Imagine that every vertex $v$ in the graph sends a runner of unit speed to every vertex of the path, starting at time $-x_v$.
After reaching the path, the runner keeps running on it until they reach the end.
We will call a runner a \emph{local leader} if they were the first one to reach the end of the path out of all the runners that entered the path at a no later position.
It is easy to see that the sequence of origins of local leaders in the order they reach the end of the path is the same as the sequence of roots of clusters into which the path is partitioned by the algorithm.
Therefore, it is enough to observe the local leaders as they reach the end of the path.
It can be shown that in any time interval $[y, y+\epsilon]$ the probability of the origin of the last local leader to reach the end of the path changing is $\Oh(\beta \epsilon)$.
Unfortunately, the entire process could take an arbitrary amount of time in the case of a path.

To apply the above reasoning to a cycle, we will 'unroll' it into a two-way infinite path.
We will set the 'finish line' at an arbitrary vertex (at position $0$) and observe the local leaders for any period of time $[y, y + L]$.

Assume the length of the cycle is $L$ and it has $l$ vertices.
Let $i \in \{0, \ldots, l - 1\}, i \in \mathbb{Z}$.
Then, the $i\cdot n + j$-th element of the sequence $s$ will intuitively be equal to the time the runner sent from the $j$-th vertex of the graph to the $i$-th vertex of the unrolled cycle reaches vertex $0$ of the unrolled cycle.
The sequence $a$ will simply label the origin of the runner relevant to the current index (and so $a_i = (i \bmod n)$).
The sequence $c$ will label the cluster to which the relevant vertex of the cycle is assigned (the origin of the first runner to reach it).
The function $f(y)$ will give the origin of the runner that reached vertex $0$ before time $y$ and entered the cycle at the earliest position.
Since only such runners will correspond to clusters, our goal will be to bound the frequency with which $f$ may change.

%\subsection{Proof Shortening}
%Let $k, n \in \mathbb{N}$ and $L \in \mathbb{R}_{+}$. For all $i \in \mathbb{Z}$ let $t_i \in \mathbb{R}$ satisfy $t_{i + k} = t_i + L$ and let $a_i \in {0, ..., n -1}$ satisfy $a_{i + k} = a_i$. 

\subsection{Periodically Decreasing Sequences}
\label{sec_perincseq}

Let $k, n \in \mathbb{N}$ and $L \in \mathbb{R}_{+}$.

Let $s_i$ be a two-way infinite sequence of real numbers indexed by $i \in \mathbb{Z}$ with the property
\begin{align*}
	\forall_i~s_{i + k} = s_i - L.
\end{align*}

Let $a_i$ be a two-way infinite sequence of integers in $\{0, \ldots, n - 1\}$ indexed by $i \in \mathbb{Z}$, periodic with period $k$, that is 
\begin{align*}
	\forall_i~a_{i + k} = a_i.
\end{align*}
We construct the sequence $c_i$ by defining
\begin{align*}
	c_i = a_j,
\end{align*}
where $j$ is the minimum $q$ that minimizes the value of $s_q$ among $q \leq i$.

We can similarly construct $f : \mathbb{R} \to \{0, \ldots, n - 1\}$ by setting for every real number $y$
\begin{align*}
	f(y) = a_j,
\end{align*}
where $j$ is the minimum $q$ that satisfies $s_q \leq y$.

\begin{fact}
	The sequence $c_i$ is periodic with period $k$.
\end{fact}

\begin{fact}
	The function $f$ is periodic with period $L$.
\end{fact}

\begin{fact}
	\label{lem:sfequi}
	For any $i \in \mathbb{Z}$ and $y \in \mathbb{R}$, the number of times the sequence $c_i, c_{i + 1}, \ldots, c_{i + k}$
	changes values is equal to the number of times $f$ changes values on the interval $[y, y + L]$. 
\end{fact}

\subsection{Random Periodically Decreasing Sequences}

Let $k, n \in \mathbb{N}, L \in \mathbb{R}_{+}$.

Let $t_i$ be a two-way infinite sequence of real numbers indexed by $i \in \mathbb{Z}$ with the property
\begin{align*}
	\forall_i~t_{i + k} = t_i - L.
\end{align*}

Let $a_i$ be a two-way infinite sequence of integers in $\{0, \ldots, n - 1\}$ indexed by $i \in \mathbb{Z}$, periodic with period $k$, that is
\begin{align*}
	\forall_i~a_{i + k} = a_i.
\end{align*}
Let $x_0, x_1, \ldots, x_{n - 1}$ be independent random variables drawn from the exponential distribution with parameter $\beta$.

We define for every $i \in Z$:
\begin{align*}
	s_i = t_i - x_{a_i}
\end{align*}
We define the function $f : \mathbb{R} \to \{0, \ldots, n - 1\}$ as in the previous section, that is
\begin{align*}
	f(y) = a_j,
\end{align*}
where $j$ is the minimum $q$ that satisfies $s_q \leq y$.

\vspace{0.3cm}

In the following lemmas, our goal will be to bound the expected number of times the value of $f$ changes on any interval.

\begin{lemma}
	\label{lem:epsconst}
	For any $y \in \mathbb{R}, \epsilon \in \mathbb{R}_{+}$, the probability that $f$ is not constant on the interval $[y, y + \epsilon]$
	is bounded by $\Oh(\beta\epsilon)$.
\end{lemma}
\Proof
	Fix $y$ and $\epsilon$.
    We condition on the value of $f(y + \epsilon)$; assume it is $k$.
    We also condition on $x_i$ for all $i \neq k$.
    Now the condition $f(y + \epsilon) = k$ is equivalent to assuming $x_k \geq c$ for some constant $c$.
    Because we have no more information about $x_k$, the conditional probability that $x_k \geq c + \epsilon$ is $1 - \Oh(\beta\epsilon)$.
    This implies the thesis.
\QED

In order to exploit \Cref{lem:epsconst} to bound the expected number of changes in $f$ we will
attempt to condition on the event $D_\epsilon$.

\begin{definition}
	Let $\epsilon \in \mathbb{R}_{+}$.
	The event $D_\epsilon$ occurs iff for all pairs $i, j \in \mathbb{Z}$ such that $a_i \neq a_j$ or $t_i \neq t_j$ it holds that
	\begin{align*}
		|s_i - s_j| > \epsilon.
	\end{align*}
\end{definition}

\begin{fact}
	\label{lem:hipdeps}
	\begin{align*}
		\lim_{\epsilon \to 0} P(D_\epsilon) = 1.
	\end{align*}
\end{fact}

Using \Cref{lem:hipdeps}, we pick an $\epsilon > 0$ that satisfies
\[
	P(D_\epsilon) \geq 1 - \min\left(\frac{1}{2}, \frac{\beta L}{k}\right)
	\enspace \text{ and } \enspace
	\frac{L}{\epsilon} \in \mathbb{N} ~.
\]
%\begin{align*}
%	P(D_\epsilon) &\geq 1 - \min\left(\frac{1}{2}, \frac{\beta L}{k}\right),\\
%	\frac{L}{\epsilon} &\in \mathbb{N}.
%\end{align*}

\begin{lemma}
	\label{lem:cepsconst}
	Assume $\epsilon$ is chosen as above.
	Conditioning on $D_\epsilon$, for any $y \in \mathbb{R}$, the probability that $f$ is not constant on the interval $[y, y + \epsilon]$
	is bounded by $\Oh(\beta\epsilon)$.
\end{lemma}
\Proof
	Because $P(D_\epsilon) \geq \frac{1}{2}$, the conditional probability is at most two times larger than in the case where we do not condition
	on $D_\epsilon$.
	The thesis follows from \Cref{lem:epsconst}.
\QED

\begin{lemma}
	\label{lem:cfewcuts}
	Assume $\epsilon$ is chosen as above.
	Conditioning on $D_\epsilon$, for any $y \in \mathbb{R}$, the expected number of times $f$ changes values
	in $[y, y + L]$ is bounded by $\Oh(\beta L)$.
\end{lemma}
\Proof
	Because we assume $D_\epsilon$, we know that $f$ can change at most once on any interval of length $\epsilon$.
	Hence it follows from \Cref{lem:cepsconst} that the expected number of time $f$ changes
	on any interval of length $\epsilon$ is bounded by $\Oh(\beta\epsilon)$.
	Because $L / \epsilon \in \mathbb{N}$, we can cover the interval $[y, y + L]$ with
	$L / \epsilon$ intervals of length $\epsilon$.
	Because of linearity of expectation, the expected number of times $f$ changes values on $[y, y + L]$ is therefore
	bounded by $\Oh(\beta L)$.
\QED

\begin{lemma}
	\label{lem:ffewcuts}
	For any $y \in \mathbb{R}$, the expected number of times $f$ changes values in $[y, y + L]$ is bounded by $\Oh(\beta L)$.
\end{lemma}
\Proof
	It follows from \Cref{lem:sfequi} that $f$ cannot change values more than $k$ times on an interval of length $L$.
	For $\epsilon$ chosen as above, we can apply this observation together with \Cref{lem:cfewcuts} to see that the expected number of changes is bounded by
	\begin{align*}
		P(D_\epsilon)\Oh(\beta L) + (1 - P(D_\epsilon))k &= \Oh(\beta L) + \frac{\beta L}{k} \cdot k =  \Oh(\beta L)
	\end{align*}
\QED

\subsection{Low-radius Decompositions}

Recall that we are considering the clustering algorithm $\textsc{Cluster-Directed}$ applied to a directed graph $G = (V, E)$.
Consider a cycle $C$ in $G$.
Assume the length of $C$ is $L$ and the number of vertices on $C$ is $l$.
Let the vertices on the cycle be $u_0, \ldots, u_{l - 1}$, in order, with $u_0$ chosen arbitrarily.

For $i \in \{0, \ldots, l - 1\}$, define $p_i$ to be the distance from $u_0$ to $u_i$ when going along the cycle.

\vspace{0.3cm}

Let $k = l\cdot n$.
We now define the two-way infinite sequence $t$ as follows, for
$z \in \mathbb{Z}, i \in \{0, \ldots, m - 1\}, j \in \{0, \ldots, n - 1\}$:
\begin{align*}
	t_{z\cdot k + i \cdot n + j} = d(v_j, u_i) - zL - p_i.
\end{align*}
We define the two-way infinite sequence $a$ for $z \in \mathbb{Z}, j \in \{0, \ldots, n - 1\}$:
\begin{align*}
	a_{z\cdot n + j} = j.
\end{align*}

\begin{fact}
	\label{lem:paths}
	Let $i \in \{0, \ldots, l - 1\}$.
	Assume $j$ is a (possibly negative) integer such that $j \leq i \cdot n + n - 1$.
	Then there exists a path from $v_{a_j}$ to $u_i$ of length $t_j + p_i$.
\end{fact}

\begin{fact}
	\label{lem:spaths}
	Let $i \in \{0, \ldots, l - 1\}, q \in \{0, \ldots, n - 1\}$.
	There exists an integer $j \leq i \cdot n + n - 1$ such that $a_j = q$ and
	\begin{align*}
		t_j + p_i = d(v_q, u_i).
	\end{align*}
\end{fact}

Recall that in $\textsc{Cluster-Directed}$ we associate with each vertex $v_i$ an independent random variable $x_i$ drawn from
the exponential distribution with parameter $\beta$.
We now define the two-way infinite sequence $s$ as

\begin{align*}
	s_i = t_i - x_{a_i}.
\end{align*}

As in Section \ref{sec_perincseq} we construct the sequence $c_i$ by defining
\begin{align*}
	c_i = a_j,
\end{align*}
where $j$ is the minimum $q$ that minimizes the value of $s_q$ among $q \leq i$.

\begin{lemma}
	\label{lem:cclusters}
	For $i \in \{0, \ldots, l - 1\}$, $c_{i \cdot n + n - 1}$ is the index of the vertex to whose cluster $u_i$ is assigned by $\textsc{Cluster-Directed}$. 
\end{lemma}
\Proof
    This follows from Facts \ref{lem:paths} and \ref{lem:spaths}.
\QED

\vspace{0.3cm}
We are now ready to prove the main theorem.

\Proofof{\Cref{lem:cycle}}
	By \Cref{lem:cclusters}, it is enough to bound the number of times $c_0, \ldots, c_k$
	changes values.
	By \Cref{lem:sfequi} this reduces to bounding the number of times the associated function $f : \mathbb{R} \to \{0, \ldots, n - 1\}$
	changes values on any interval of length $L$.
	This is shown to be $\Oh(\beta L)$ in expectation in \Cref{lem:ffewcuts}.
\QED

\Proofof{\Cref{thm:ddecomp}}
    First note that with high probability $\max(x_1, \ldots, x_n) \leq r$, and so the radius of the computed clusters is at most $r$.
    To bound the number of cut edges, it is enough to note that by \Cref{thm:definitions}, we can find a set of simple cycles $C_1, \ldots, C_k$ such that their total volume is at most $\bal(G)\vol(G)$ and has weight at least $w(e)$ on every edge $e \in E$.
    The thesis follows by applying \Cref{lem:cycle}.
\QED

%% file: lowstretch_proofs.tex
\section{Missing proofs from \Cref{sec:lowstretch}}
\label{sec:lowstretch_proofs}

\begin{definition}
    We define the additive imbalance $\abal(G)$ of a directed graph $G$ as the minimum $\iota$ such that it is possible to add edges of total weight $\iota$ to $G$ to make it Eulerian.
\end{definition}

In order to make the running time of our algorithm independent of the diameter of the graph, we will attempt to \emph{collapse} very short edges in the upper levels of the recursion, that is, contract their endpoints into a single vertex.
This is similar to the scheme proposed in \cite{CohenMPPX14,CohenKMPPRX14}.
However, this operation is not always feasible in directed graphs; thus, we will only perform the contraction if both endpoints of the edge can reach each other by following only very short edges.

\begin{definition}
    Let $G = (V, E, w, l)$ be a directed graph and $x_L, x_R \in \mathbb{R}$ be such that $0 < x_L < x_R$.
    We construct $G$ \emph{collapsed} to $[x_L, x_R]$ by:
    \begin{itemize}
        \item merging any vertices that can reach each other while following only arcs of length at most $x_L$, and
        \item reducing the length of all arcs longer than $x_R$ to $x_R$.
    \end{itemize}
\end{definition}

\begin{figure}[ht]
\noindent
\centering
\fbox{
\begin{minipage}{6in}
    \noindent $T = \textsc{Find-Arborescence} (G, s)$, where $G = (V, E, l)$ is a directed graph and $s \in V$ is such that all vertices in $G$ are reachable from $s$.
\begin{enumerate}
    \item If $n = 1$, return a single-vertex graph.
    \item Let $r := \max_{v \in V} d_G(s, v)$.
    \item Let $r' := r / (c \cdot \log n)$.
    \item Let $G'$ be the graph $G$ collapsed to $[r' / n, 2r']$. Let $s'$ be the vertex in $G'$ corresponding to $s$.
    \item Let $V'_1, V'_2, \ldots, V'_k := \textsc{Cluster-Directed-Rooted}(G', s', r')$.
    \item Expand the clusters $V'_1, \ldots, V'_k$ back into $G$, obtaining $V_1, \ldots, V_k$.
    \item Let $G_i$ be the graph induced by $V_i$, for $i = 1, \ldots k$, and $u_i$ denote the center of cluster $V_i$ (with $u_1 = s_1$).
        \item Let $
                T' := \bigcup_{i=1}^k \textsc{Find-Arborescence}(G_i, u_i).
            $
        \item Let $T$ be $T'$ with the arcs $(s, u_i)$ of length $d_G(s, u_i)$ added for each $i = 2, \ldots, k$.
    \item Return $T$.
\end{enumerate}
\end{minipage}
}
\caption{The low-stretch arborescence finding algorithm.}
\label{fig:lowstretch}
\end{figure}

\begin{figure}[ht]
\noindent
\centering
\fbox{
\begin{minipage}{6in}
    \noindent $(V_1, V_2, \ldots) = \textsc{Cluster-Directed-Rooted} (G, s, r)$, where $G = (V, E, l)$ is a directed graph, $s \in V$ and $r > 0$.
\begin{enumerate}
    \item Choose $r'$ uniformly at random from $[0, r]$.
    \item Let $V_1$ be the set of vertices at distance at most $r'$ from $s$.
    \item Let $G'$ be the induced graph on $V - V_1$.
    \item Let $V_2, V_3, \ldots V_k := \textsc{Cluster-Directed}(G', r)$.
    \item Return $V_1, V_2, \ldots, V_k$.
\end{enumerate}
\end{minipage}
}
\caption{The decomposition algorithm with a specified root.}
\label{fig:ddecomp_root}
\end{figure}

\begin{lemma}
\label{lem:ddecomp_imbalance}
    Let $G = (V, E, w, l), s \in V, r > 0$.
    Let $V_1, \ldots, V_k = \textsc{Cluster-Directed-Rooted}(G, s, r)$.
    Then:
    \begin{itemize}
        \item each cluster $V_i$ has radius at most $r$,
        \item the cluster $V_1$ containing $s$ has radius at most $r$ from $s$,
        \item the expected total weight of edges going between different clusters is $\Oh(\vol(G) \log n / r + \abal(G) \log n)$, and
        \item the expected total additive imbalance of the clusters is $\Oh(\vol(G) \log n / r + \abal(G) \log n)$.
    \end{itemize}
    Moreover, the algorithm works in expected linear time.
\end{lemma}
\Proof
First, note that the expected total weight of edges between $V_1$ and $V - V_1$ is $\Oh(\vol(G) / r)$.
Hence the expected additive imbalances of the cluster on $V_1$ and that of $G'$ are both $\Oh(\abal(G) + \vol(G) / r)$.

By the definition of additive imbalance, we can add edges of expected total weight $\Oh(\abal(G) + \vol(G) / r)$ to $G'$ to make it Eulerian.
    We obtain the graph $G''$ by adding such edges, each with length $2r$.
    The expected volume of $G''$ is $\Oh(\vol(G)) + \Oh(\abal(G) + \vol(G) / r)\cdot 2r = \Oh(\vol(G) + \abal(G) \cdot r)$.
    Now by \Cref{thm:ddecomp} we can partition $G''$ into clusters of radius at most $r$, with the expected total weight of edges going between clusters $\Oh(\vol(G'') \log n / r) = \Oh(\vol(G) \log n / r + \abal(G) \log n)$.
    Note that if we remove the added edges, the radii of these clusters cannot change, as the edges have length greater than $r$; at the same time, their total additive imbalance can increase by at most $\Oh(\abal(G) + \vol(G) / r)$ in expectation.
    To complete the analysis, observe that in fact the edges added in the above reasoning are ignored by the decomposition algorithm.
    Hence, they are only necessary for the analysis.
\QED

\begin{lemma}
    \label{lem:collapse}
    Let $G = (V, E, w, l)$ be a directed Eulerian graph and $x_L, x_R \in \mathbb{R}$ be such that $0 < x_L < x_R$.
    Let $G' = (V', E', w', l')$ be $G$ collapsed to $[x_L, x_R]$.
    Then $\vol(G')$ is at most
    \begin{align*}
        2\cdot\sum_{e\in E : l(e) > x_L / n} w(e)\min(l(e), x_R).
    \end{align*}
\end{lemma}
\Proof
Since $G'$ is Eulerian, it can be represented as a sum of simple cycles of uniform weight.
Consider any such decomposition and take any cycle $C$ in it.
Then $C$ must contain an edge of length at least $x_L$, and it contains at most $n$ edges of length not exceeding $x_L / n$.
Hence, the length of $C$ is at most two times greater than the sum of its edge lengths greater than $x_L / n$.
Summing over all the cycles yields the desired bound.
\QED

\Proofof{\Cref{thm:lowstretch}}
    First, note that by \Cref{thm:definitions} the edge weights in $G$ can be increased to obtain an Eulerian graph with volume at most $\bal(G)\vol(G)$.
    Since the algorithm is oblivious to weights, it is enough to consider Eulerian graphs in the proof; from now on we assume $\bal(G) = 1$.

    Properties $1$ and $2$ are easy to verify.
    Assume the constants hidden in the big-oh notation in \Cref{lem:ddecomp_imbalance} are bounded by $c_0$.
    We set $c := 2c_0 + 4$.

    Consider the $i$-th level (numbering from $0$) of the tree of recursive calls of $\textsc{Find-Arborescence}(G, s)$.
        Let $r_i = r / (c \log n)^i$.
        It can easily be shown by induction that the radii of the graphs in the $i$-th level are at most $r_i$, and the radii of the returned arborescences are at most $2r_i$, since $c \geq 4$.
        Let $\nu_i$ be the total volume of the \emph{collapsed} graphs at level $i$.

    By \Cref{lem:ddecomp_imbalance} the additive imbalance of the graphs in the $i$-th level can be bounded by
    \begin{align*}
          &(c_0 \log n)^i \cdot \nu_0 / r_1\\
        + &(c_0 \log n)^{i - 1} \cdot \nu_1 / r_2\\
        + &(c_0 \log n)^{i - 2} \cdot \nu_2 / r_3\\
        + &\ldots\\
        + &(c_0 \log n)^1 \cdot \nu_{i-1} / r_i.
    \end{align*}
    Since $c > 2c_0$, the above sum is bounded by
    \begin{align*}
        (c \log n)^{i+1} \sum_{j < i} \left(\nu_j / 2^{i-j}\right).
    \end{align*}
    Hence, the total weight of edges cut at level $i$ is at most
    \begin{align*}
        (c_0 \log n) \left(\nu_i / r_{i+1} + (c \log n)^{i+1} \sum_{j < i} \left(\nu_j / 2^{i-j}\right)\right)
        &\leq
        (c \log n)^{i+2}/2 \cdot \sum_{j \leq i} \left(\nu_j / 2^{i-j}\right).
    \end{align*}
    Since the radius of the arborescence returned at level $i$ is at most $2r_i$, we have that the total stretch incurred at level $i$ is at most \begin{align*}
        2r_i \cdot (c \log n)^{i+2}/2 \cdot \sum_{j \leq i} \left(\nu_j / 2^{i-j}\right).
        \leq
        (c \log n)^2 \cdot \sum_{j \leq i} \left(\nu_j / 2^{i-j}\right).
    \end{align*}
    Hence the total stretch is at most
    \begin{align*}
        (c \log n)^2 \cdot \sum_i \sum_{j \leq i} \left(\nu_j / 2^{i-j}\right)
        &= (c \log n)^2 \cdot \sum_j \left(\nu_j 2^j \sum_{i\geq j} 2^{-i}\right)\\
        &\leq 2(c\log n)^2 \cdot \sum_j \nu_j.
    \end{align*}
    Observe that all the collapsed graphs at level $j$ are subgraphs of $G$ collapsed to $[r_{j+1}/n, 2r_{j+1}]$.
    Hence, by \Cref{lem:collapse}, we have
    \begin{align*}
        \nu_j \leq 2\cdot \sum_{e \in E : l(e) > r_{j+1}/n^2} w(e)\min(l(e), 2r_{j+1}).
    \end{align*}
    Hence
    \begin{align*}
        \sum_j \nu_j &\leq 2\cdot \sum_{e \in E} \sum_{j : r_{j+1} < l(e) \cdot n^2} w(e)\min(l(e), 2r_{j+1})\\
                     &= \Oh(\vol(G) \log n / \log \log n).
    \end{align*}
    Combining this with the previous bound yields the thesis.
\QED

%% file: racke_proofs.tex
\section{Missing proofs from \Cref{sec:racke}}

\Proofof{\Cref{lem:racke}}
It is enough to note that in step 2c) of \Cref{fig:racke}, with high probability, $T_k'$ is a tree with total stretch $\Oh(\bal(G) \log^3 n / \log \log n)$ in $G_k'$, where $T_k'$ and $G_k'$ are undirected counterparts of $T_k$ and $G_k$, respectively.
Hence, the analysis of \cite{Racke08} can be applied to complete the proof.
\QED

%% file: lower_proofs.tex
\section{Missing proofs from \Cref{sec:lower}}
\label{sec:lower_proofs}

\Proofof{\Cref{thm:lowerbound}}
Let $k \geq 1$.
Let $G$ be a directed graph on the vertex set
\begin{align*}
    V = S \cup T \cup \{s\} \cup \{t\}, \mbox{ where $|S| = |T| = k$}
\end{align*}
and edge set
\begin{align*}
    E &= S \times T \mbox{ with weight $1$}\\
      &\cup S \times \{s\} \mbox{ with weight $k$}\\
      &\cup \{(s, t)\} \mbox{ with weight $k$}\\
      &\cup \{t\} \times T \mbox{ with weight $k$}.
\end{align*}
Assume some oblivious routing $\mathcal{A}$ achieves competitive ratio $c$ on $G$.
Let $u \in S$ and $v \in T$.
The optimal congestion for the unit flow from $u$ to $v$ is at most $1 / k$, which can be achieved by routing the flow through $s$ and $t$.
Therefore, $\mathcal{A}$ must achieve congestion at most $c / k$, hence putting at least $1 - c / k$ units of flow on the edge $(s, t)$.

The optimal congestion for the multicommodity flow with unit demand between every pair in $S \times T$ is clearly at most $1$.
Simultaneously, by the above argument, $\mathcal{A}$ must put at least $k(k - c)$ flow on the edge $(s, t)$.
Hence we have $c \geq k - c$, implying $c \geq k/2$.
As $n = 2k + 2$ we have $c = \Omega(n)$.
\QED
\Proofof{\Cref{thm:lowerboundbalanced}}
Let $n \geq 2$.
Let $G$ be a directed graph on the vertex set
\begin{align*}
    V = \{v_1, \ldots, v_n\}
\end{align*}
and edge set
\begin{align*}
    E &= C_1 \cup C_2,\mbox{ where}\\
    C_1 &= \{(v_1, v_2), (v_2, v_3), \ldots, (v_{n-1}, v_n), (v_n, v_1)\}\mbox{ with weight $1$, and}\\
    C_2 &= \{(v_n, v_{n-1}), (v_{n-1}, v_{n-2}), \ldots, (v_2, v_1), (v_1, v_n)\}\mbox{ with weight $\sqrt{n}$}.
\end{align*}
Note that $G$ is Eulerian.
Assume some oblivious routing $\mathcal{A}$ achieves competitive ratio $c$ on $G$.
Let $i < n$.
The optimal congestion for the unit flow from $v_i$ to $v_{i+1}$ is at most $1 / \sqrt{n}$, which can be achieved by routing the flow through $C_2$.
Therefore, $\mathcal{A}$ must achieve congestion at most $c / \sqrt{n}$, hence putting at least $1 - c / \sqrt{n}$ units of flow on the edge $(v_n, 1)$.

The optimal congestion for the multicommodity flow with unit demand between every such pair $(v_i, v_{i+1})$ is clearly at most $1$.
Simultaneously, by the above argument, $\mathcal{A}$ must put at least $(n-1)(1 - c / \sqrt{n})$ flow on the edge $(v_n, 1)$.
Hence we have $c \geq (n-1) / \sqrt{n} - c$, implying $2c \geq \sqrt{n - 1}$.
Therefore $c = \Omega(\sqrt{n})$.

%% file: maxflow_proofs.tex
\section{Missing proofs from \Cref{sec:maxflow}}

\Proofof{\Cref{lem:maxflow-correct}}

We have
\begin{align*}
    \grad \mu(f) = -2\alpha \mb^\top v.
\end{align*}

Therefore
\begin{align*}
    2\alpha\cdot\|\mc\max(\mb^\top v, 0)\|_1 \leq 1 + \frac{\epsilon}{4}.
\end{align*}

It also holds that
\begin{align*}
    \normInf{\mc^{-1}f} + 2\alpha v^\top(b - \mb f)
    &= \normInf{\mc^{-1}f} + p^T x\\
    &\geq \phi(f) - 4\ln n\\
    &\geq \left(1 - \frac{\epsilon}{4}\right) \phi(f).
\end{align*}

Simultaneously, we have
\begin{align*}
    \frac{\epsilon}{4} \phi(f) \geq \frac{\epsilon}{4} &\geq \normInf{\mc^{-1}f} + \grad \mu(f)^T f\\
                               &= \normInf{\mc^{-1}f} - 2\alpha f^T\mb^\top v.
\end{align*}

Hence
\begin{align*}
    2\alpha v^Tb \geq \left(1 - \frac{\epsilon}{2}\right) \phi(f),
\end{align*}
and so
\begin{align*}
    \frac{b^T v}{\|\mc\max(\mb^\top v, 0)\|_1} \geq \frac{\phi(f)}{1 + \epsilon}.
\end{align*}
\QED

\Proofof{\Cref{lem:maxflow-runtime}}
    Let us call the iterations between each scaling in step 2a) a \emph{phase}.
    Since the initial scaling gives us the correct scale to within factor $1 + \epsilon_0$, we will scale at most $\Oh(\log (1 + \epsilon_0))$ times.
    Moreover, if $\epsilon_0 < 1/10$, step 2a) will never be executed.

    If step 2b) is about to be executed, then
    \begin{align*}
        \phi(f + \delta s) &\leq \phi(f) + \delta + \delta \grad \mu(f)^\top s + 2\alpha^2 \delta^2\\
                        &\leq \phi(f) - \frac{\epsilon\delta}{4} + 2\alpha^2 \delta^2.
    \end{align*}

    If step 2c) is about to be executed, then
    \begin{align*}
        \phi(f - \delta f) &\leq \phi(f) - \delta \normInf{\mc^{-1} f} - \delta \grad \mu(f)^\top f + 2\alpha^2 \delta^2\\
                        &\leq \phi(f) - \frac{\epsilon\delta}{4} + 2\alpha^2 \delta^2.
    \end{align*}

    In both cases we have
    \begin{align*}
        \frac{\epsilon\delta}{4} - 2\alpha^2\delta^2
            &\geq \frac{\epsilon^2}{40\alpha^2} - \frac{\epsilon^2}{50 \alpha^2}\\
            &=  \frac{\epsilon^2}{200\alpha^2}.
    \end{align*}

    Hence each iteration of steps 2b) and 2c) decreases $\phi(f)$ by at least $\Omega(\epsilon^2\alpha^{-2})$.
    
    For $\epsilon_0 \geq 1/10$, every scaling in step 2a) increases $\phi(f)$ by at most $\epsilon^{-1} \ln n$.
    Hence, for such $\epsilon_0$ there can be at most $\Oht(\log(1 + \epsilon_0) \alpha^2 \epsilon^{-3})$ iterations in total.

    For $\epsilon_0 < 1/10$, step 2a) will never be executed.
    Moreover, the $\phi(f)$ after the initial scaling must be at most $OPT_b + \Oht(\epsilon_0 \epsilon^{-1})$.
    Hence steps 2b) and 2c) can be executed at most $\Oht(\epsilon_0 \alpha^2 \epsilon^{-3}) = \Oht(\log(1 + \epsilon_0) \alpha^2 \epsilon^{-3})$ times.

\QED

\Proofof{\Cref{lem:fastroute}}
    As $\phi(\vec0) = \Oht(OPT_b \alpha)$, step 1. works in $\Oht(m\alpha^2)$ time by \Cref{lem:maxflow-runtime}.

    Now note that we can apply \Cref{thm:grad-desc-progress} to $\psi'(x) + g(x)$ with $\psi'(x) = \psi(\mc x), g(x) = \mu(\mc x), L = \alpha^2, D = 50 \epsilon^{-1} \ln n$.
    This yields that $\phi(f_K) \leq (1 + \Oht(\epsilon)) OPT_b$.
    Hence by \Cref{lem:maxflow-runtime}, step 4. runs in $\Oht(m\alpha^2\epsilon^{-2})$ time.

    The only remaining thing to show is that we can solve the optimization problem in step 3. in $\Oht(m)$ time.
    It can be reformulated by introducing an auxiliary variable $z$:

    \begin{equation*}
        \begin{aligned}
            & \underset{f, z}{\text{minimize}}
            & & \grad \mu(f_k)^\top f + \frac{1}{2}\alpha^2z^2 + \psi(f) \\
            & \text{subject to}
            & & \normInf{\mc^{-1}(f - f_k)} \leq z.
        \end{aligned}
    \end{equation*}

    For a fixed $z$, the problem can easily be solved in $\Oh(m \log m)$ time by sorting.
    Hence we can employ ternary search over $z$ to achieve $\Oht(m)$ runtime.
\QED

\Proofof{\Cref{lem:isbalanced}}
Construct $G'$ by adding the reverse of $G$ multiplied by $\frac{1}{4\alpha}$ to $G$.
Note that $\bal(G') \leq 4\alpha$.
Let $b'$ be the residual degrees in $G'$.
Now by \Cref{thm:maxflow} we can compute a $2$-overestimate $c'$ to the minimum congestion to route $-b'$ in $G'$, in time $\Oht(m \alpha^2)$.
Note that we have
\begin{align*}
    \bal(G') \leq c' - 1 \leq 2\bal(G') \leq 2\bal(G).
\end{align*}
Hence if $c' - 1 > 2\alpha$ we can conclude that $\bal(G) > \alpha$ and return the corresponding cut.

Otherwise, we must have $\bal(G) \leq 2\alpha$.
Hence we can compute a $(1+\epsilon)$-overestimate $c$ to the minimum congestion to route $-b$ in $G$, in time $\Oht(m\alpha^2\epsilon^{-2})$, where $b$ are the residual degrees in $G$.
Now we have
\begin{align*}
    \bal(G) \leq c - 1 \leq (1+\epsilon)\bal(G),
\end{align*}
and so if $c - 1 \leq \alpha$ then $\bal(G) \leq \alpha$, and otherwise we can return a cut proving that $\bal(G) > (1-\epsilon)\alpha$.
\QED

\Proofof{\Cref{lem:sparsest}}
    First, we can use \Cref{lem:isbalanced} to check whether $\bal(G) \leq \phi^{-1}$.
    If it is not, we can return the smaller direction of the imbalanced cut as the result.
    Otherwise, we use can apply the cut-matching game algorithm given by Louis \cite{Louis10} for $\phi' = \frac{n\phi}{4}$ \footnote{The rescaling by $n$ is used due to a slightly different definition of sparsity in \cite{Louis10}.} and reduce the problem to a sequence of $\Oht(1)$ maximum flow queries.
    Each of the queries fixes some $S \subseteq V$ with $|S| = n/2$ and asks for a flow in $G$ with demands $-\phi'$ on $S$ and $\phi'$ on $V \setminus S$.
    We can compute the $2$-approximate minimum congestion flow for such a query.
    If the returned congestion is at most $1$, we return the flow.
    Otherwise, we have a set $T \subseteq V$ which achieves
    \begin{align*}
        b_T / w(T, V \setminus T) &\geq \frac{1}{2},\\
        w(T, V - T) &\leq 2b_T\\
                    &\leq 2\phi' \min(|T|, |V\setminus T|)\\
                    &\leq \frac{n}{2}\phi \min(|T|, |V\setminus T|)\\
                    &\leq \phi |T|\cdot|V\setminus T|.
    \end{align*}
\QED

%% file: proximal_proofs.tex
\section{Missing proofs from \Cref{sec:proximal}}

We break the proof into 2 parts, first we prove a lemma about the
progress of each gradient step and then we use this to prove the theorem. Let $X_{*}$ be the set of all optimal solutions to $\min_{x} f(x)$.

\begin{lemma}[Gradient Descent Progress]
 \label{lem:progress} For all $k\geq0$ we have that for all $x_{*}\in X_{*}$
\[
f(x_{k+1}) \leq f(x_{k})-\min\left\{ \frac{1}{2L}\left(\frac{\epsilon_{k}}{\norm{x_{k}-x_{*}}}\right)^{2},\frac{f(x_{k})-f(x_{*})}{2}\right\} 
\]
\end{lemma}
\Proof
By the smoothness of $g$ we know that for all $x\in\R^{n}$ we have
\[
f(x)\leq g(x_{k})+\innerProduct{\nabla g(x_{k})}{x-x_{k}}+\frac{L}{2}\norm{x-x_{k}}^{2}+\psi(x)\,.
\]
By definition of $x_{k+1}$ we then have that
\[
f(x_{k+1})\leq\min_{x} \Big(g(x_{k})+\innerProduct{\nabla g(x_{k})}{x-x_{k}}+\frac{L}{2}\norm{x-x_{k}}^{2}+\psi(x) \Big)\,.
\]
Now it follows from the convexity of $g$ that 
\[
g(x)\geq g(x_{k})+\innerProduct{\nabla g(x_{k})}{x-x_{k}},
\]
and combining these yields that
\begin{equation}
f(x_{k+1})\leq\min_{x\in\R^{n}} \Big(f(x)+\frac{L}{2}\norm{x-x_{k}}^{2}\Big) \label{eq:lem:progress:1}
\end{equation}
Since $f$ is convex, for all $\alpha\in[0,1]$ and $x_{*}\in X_{*}$ we have
\[
f(\alpha x_{*}+(1-\alpha)x_{k})\leq\alpha f(x_{*})+(1-\alpha)f(x_{k}) = f(x_{k})-\alpha(f(x_{k})-f(x_{*})).
\]
Consequently 
\[
\min_{x\in\R^{n}} \Big(f(x)+\frac{L}{2}\norm{x-x_{k}} \Big) \leq\min_{\alpha\in[0,1]} \Big(f(x_{k})-\alpha (f(x_{k})-f(x_{*}))+\frac{L\alpha^{2}}{2}\norm{x_{k}-x_{*}}^{2}\Big).
\]
By taking the derivative with respect to $\alpha$ of the expression on the right hand side above and setting it to zero, we see that the optimal $\alpha$ satisfies
\[
- (f(x_{k})-f(x_{*}))+\alpha L\norm{x_{k}-x_{*}}^{2}=0
\]
and thus using $\alpha=\min\left\{ \frac{f(x_{k})-f(x_{*})}{L\norm{y-x_{*}}^{2}},1\right\} $
yields the result, since $f(x_{k})-f(x_{*})\geq 0$, and when $f(x_{k})-f(x_{*})\geq L\norm{x_{k}-x_{*}}^{2}$, we have
\[
\min_{x\in\R^{n}} \Big(f(x)+\frac{L}{2}\norm{x_{k}-x}^{2} \Big)\leq f(x_{*})+\frac{L}{2}\norm{x_{k}-x_{*}} \leq f(x_{*})+\frac{f(x_{k})-f(x_{*})}{2}\,.
\]
\QED
Using the lemma, we can complete the proof of the theorem as follows.

\Proofof{\Cref{thm:grad-desc-progress}}
 By \Cref{lem:progress} we have that $\epsilon_{k+1}\leq\epsilon_{k}$
for all $k$ and 
\[
\epsilon_{k+1}\leq\max\left\{ \epsilon_{k}-\frac{1}{2L}\left(\frac{\epsilon_{k}}{D}\right)^{2},\frac{\epsilon_{k}}{2}\right\} 
\]
Consequently for $k\geq1$ such that $\epsilon_{k}-\frac{1}{2L}\left(\frac{\epsilon_{k}}{D}\right)^{2}\geq\frac{\epsilon_{k}}{2}$
we have 
\[
\frac{1}{\epsilon_{k}}-\frac{1}{\epsilon_{k+1}}\leq\frac{\epsilon_{k+1}-\epsilon_{k}}{\epsilon_{k}\epsilon_{k+1}}\leq-\frac{1}{2L}\cdot\frac{1}{D^{2}}\cdot\frac{\epsilon_{k}}{\epsilon_{k+1}}\leq-\frac{1}{2LD^{2}}
\]
Summing yields that 
\[
\frac{1}{\epsilon_{1}}-\frac{1}{\epsilon_{k}}\leq-\frac{N_{k}}{2LD^{2}}
\]
where $N_{k}$ is the number of steps $k\geq1$ for which $\epsilon_{k}-\frac{1}{2L}\left(\frac{\epsilon_{k}}{D}\right)^{2}\geq\frac{\epsilon_{k}}{2}$
. Furthermore, clearly by \ref{lem:progress} we have that 
\[
\epsilon_{1}\leq\frac{L}{2}D^{2}
\]
and thus
\[
\epsilon_{k}\leq\frac{2LD^{2}}{N_{k}+4}
\]

On the other hand we have that 
\[
\epsilon_{k+1}\leq\left(\frac{1}{2}\right)^{k-1-N_{k}}
\]
and noting that either $N_{k}\geq\lfloor\frac{k-1}{2}\rfloor$ or
$k-1-N_{k}\geq\lfloor\frac{k-1}{2}\rfloor$ then yields the result.%\footnote{This could probably possibly have been done more cleverly, but this seems clean enough.}
\QED